\documentclass[prc,aps,superscriptaddress,twocolumn,showpacs,nofootinbib]{revtex4}

\usepackage[colorlinks,linkcolor=blue,anchorcolor=blue,citecolor=blue]{hyperref}
\usepackage[usenames,dvipsnames]{color}
\usepackage{graphicx,amsmath,amssymb,bm,multirow}
\usepackage{amsthm}
\usepackage{amsfonts}

\usepackage{dcolumn}
\usepackage{bm}
\usepackage{CJK}

\usepackage[usenames,dvipsnames]{color}
\begin{document} 

\title{Global dynamical correlation energies in covariant density functional theory: cranking approximation}

\author{Q. S. Zhang}
\affiliation{School of Physical Science and Technology, Southwest University, Chongqing 400715, China}

\author{Z. M. Niu}
\affiliation{School of Physics and Material Science, Anhui University, Hefei 230039, China}

\author{Z. P. Li}\thanks{zpliphy@swu.edu.cn}
\affiliation{School of Physical Science and Technology, Southwest University, Chongqing 400715, China}

\author{J. M. Yao}\thanks{jmyao@swu.edu.cn}
\affiliation{School of Physical Science and Technology, Southwest University, Chongqing 400715, China}
\affiliation{Department of Physics, Tohoku University, Sendai 980-8578, Japan}

\author{J. Meng}\thanks{mengj@pku.edu.cn}
\affiliation{State Key Laboratory of Nuclear Physics and Technology, School of Physics,
Peking University, Beijing 100871, China}
\affiliation{School of Physics and Nuclear Energy Engineering, Beihang University, Beijing 100191, China}
\affiliation{Department of Physics, University of Stellenbosch, Stellenbosch, South Africa}


\begin{abstract}
The global dynamical correlation energies for 575 even-even nuclei with proton numbers ranging from $Z=8$ to $Z=108$ calculated with the covariant density functional theory using the PC-PK1 parametrization are presented. The dynamical correlation energies include the rotational correction energies obtained with the cranking approximation and the quadrupole vibrational correction energies. The systematic behavior of the present correlation energies is in good agreement with that obtained from the projected generator coordinate method using the SLy4 Skyrme force although our values are systematically smaller. After including the dynamical correlation energies, the root-mean-square deviation predicted by the PC-PK1 for the 575 even-even nuclei masses is reduced from 2.58 MeV to 1.24 MeV.
\end{abstract}

\maketitle


With the advances of radioactive ion beams, our knowledge of nuclear physics has been extended from the stable nuclei near the stability line to the unstable nuclei far from the stability line$-$so-called exotic nuclei~\cite{tanihata85,Mueller93,Tanihata95}, in which a number of entirely unexpected features and novel aspects of nuclear structure appear, e.g. the halo phenomenon~\cite{meng96,meng98}, the erosion of traditional shell gaps~\cite{Sorlin08}, the occurrence of new ones~\cite{Ozawa00}, and the shape of the halo ¡°decoupled¡± from the shape of the ¡°core¡± nucleus~\cite{Zhou10}.
The exotic nuclei play important roles not only in nuclear physics but also in astrophysics since their properties, including masses, beta-decay rates, level densities, and fission barriers etc.,  are crucial to stellar nucleosynthesis processes such as the $r$-process~\cite{burbidge57,Sun08,Niu09,LiZ12,Xu13,zhang12,Niu13}. Due to the short lives of the exotic nuclei, their properties are difficult to measure and only a few data are available. Therefore, a microscopic nuclear model which is reliable for the description of both stable and exotic nuclei is required to deepen our knowledge of nuclear structure and to understand the nucleosynthesis in the stars.

The nuclear mass is one of the fundamental properties which reflects the binding energy
of nucleons and, thus, all interactions among them. The correct and high-precision description of nuclear binding energies is one of the major challenges for nuclear models. In the past decade, tremendous progress has been made and a number of global nuclear mass models have been proposed.

The finite-range droplet model (FRDM)~\cite{moller95} is a success macroscopic-microscopic mass model, in which the macroscopic part is described by the modified Bethe-Weizs\"{a}cker's formula~\cite{Bethe36}, and the microscopic part is composed of shell correction and pairing correlation. Recently, a semiempirical nuclear mass formula--the Weizs\"{a}cker-Skyrme (WS) model with less parameters has been proposed, which is based on the FRDM and the Skyrme energy-density functional~\cite{Wang10,Liu11}. In the WS model, the root-mean-square (rms) deviation with respect to 2149 known nuclear masses is 0.336 MeV. Other successful global nuclear mass formulas can be found in Refs.~\cite{DZ95,Koura05}.

Compared with the above mentioned mass models, the self-consistent mean-field implementation of density functional theory (DFT) stands out as a unique microscopic model that is able to describe nuclei from light to heavy mass regions in a unified way. A beyond mean-field calculation demonstrated that the dynamical correlation energies associated with the quadrupole shape degrees of freedom could improve the DFT description of binding energies and nucleon separation energies further~\cite{Bend05,Bend06}. In the framework of DFT with the Skyrme force, a nuclear mass model has been developed from the Hartree-Fock-Bogoliubov (HFB) approach and has proven its capacity to reproduce the 2149 experimental masses with a rms deviation of 0.581 MeV by considering the Wigner terms phenomenologically and the dynamical correlation energies with the cranking approximation~\cite{Gori09a}. Using the newly adjusted Gogny force D1M, the HFB approach reproduces the nuclear masses with an accuracy comparable with the mass model of the Skyrme force after including the dynamical correlation energies explicitly within a five-dimensional collective Hamiltonian (5DCH) approach~\cite{Gori10}. These studies have demonstrated the importance of the dynamical correlation effects beyond the mean-field approximation in improving the DFT description of nuclear masses~\cite{Gori09b,Bend05,Bend06,Gori10}.

The covariant density functional theory (CDFT), which relies on the basic ideas of the effective field theory and the DFT, has achieved great success in the description of ground state properties of both spherical and deformed nuclei all over the nuclear chart~\cite{Reinhard89,Ring96,Vretenar05,Meng06}. In the CDFT with a meson-exchange effective Lagrangian,  Hirata {\em et al.} constructed the first mass table without the pairing correlations for 2174 even-even nuclei with $8 \leqslant Z \leqslant 120$~\cite{Hirata97}. Later on, Lalazissis {\em et al.} developed a mass table for 1315 even-even nuclei with $10 \leqslant Z \leqslant 98$ by including the pairing correlations with a constant-gap BCS method~\cite{Lalazissis99}. In Ref.~\cite{Geng05}, Geng {\em et al.} carried out a global study of the ground state properties of more than 7000 nuclei using a state-dependent BCS method with a zero-range force. The obtained rms deviation with respect to the data of around 2000 measured nuclear masses is about 2.25 MeV. However, here it should be noted that these covariant density functionals were not established for mass models and their parameters in the effective Lagrangian were optimized for a few selected nuclear mass data only. Furthermore the dynamical correlation energies and the Wigner terms were not considered.

In recent years, the CDFT with a point-coupling  effective Lagrangian has attracted a lot of attention. It shows great advantages in its extension for describing the nuclear low-lying excited states by implementing the projections and generator coordinate method (GCM)~\cite{Niks06,Yao08,Yao09,Yao10}, the 5DCH~\cite{Niks09,Li09}, or the quasiparticle random phase approximation (QRPA) method~\cite{Niu2013PRC}. In Ref.~\cite{Yao11}, the dynamical correlation energies in Mg isotopes were calculated with the exact angular-momentum-projected (AMP) and GCM. The full dynamical quadrupole correlation energy turns out to be dominated by the rotational correction which is sensitive to nuclear shape and shell structure. Similar phenomenon has also been observed previously in a beyond nonrelativistic mean-field calculation~\cite{Bend06}. Recently, a new parameterization of the point-coupling effective Lagrangian, PC-PK1, was proposed by fitting the binding energies of 60 spherical nuclei and the charge radii of 17 spherical nuclei from O to Pb isotopes~\cite{Zhao10}. The success of the PC-PK1 has been illustrated through the description of infinite nuclear matter and finite nuclei including the ground state and low-lying excited states~\cite{Zhao10,Zhao11a,Zhao11b,Xiang12,Li12,hua12}. Moreover, by including the rotational correction, the newly-measured masses for the neutron-rich nuclei from Sn to Pa have been reproduced within 1 MeV~\cite{Zhao12} and the rms deviation by the PC-PK1 force for about 2000 measured nuclear masses is reduced from 2.25 MeV~\cite{Geng05} to 1.42 MeV~\cite{Meng13}. Most recently, masses of 402 nuclei ranging from O to Ti isotopes have been investigated with the spherical relativistic continuum Hartree-Bogoliubov (RCHB) theory~\cite{meng96} using the density functional PC-PK1. The rms deviation for the 234 nuclei with mass measured turns out to be 2.23 MeV~\cite{Qu13}.

In view of these facts, it is very interesting to investigate the global dynamical correlation energies associated with quadrupole deformations within the CDFT for the nuclei with their masses measured. The aim of this work is to provide the dynamical correlation energies for the 575 even-even nuclei with $8\leq Z\leq108$ and examine the influence of these correlation energies on the predicted masses.


A detailed description of the CDFT using the point-coupling effective Lagrangian has been given previously in Refs.~\cite{Niks06,Yao09,Yao10,Niks09,Zhao10,Burv02} and references therein. Therefore here only some numerical details are presented.

In the CDFT calculations, Dirac equation is solved by expanding in a set of eigenfunctions of an axially deformed harmonic oscillator in cylindrical coordinates with 14 major shells for nuclei with $Z<60$ and with 18 major shells for the heavier ones. It turns out that the numbers of oscillator shells are sufficient to give a reasonable good description for the binding energy. An axial deformation parameter $\beta_2$ is defined by the mass quadrupole moment as $\beta_2=\dfrac{4\pi}{3AR^2}\langle Q_{20}\rangle$ with $R=1.2A^{1/3}$ and $\langle Q_{20}\rangle=\sqrt{\dfrac{5}{16\pi}}\langle 3z^2-r^2\rangle$. The pairing correlation is taken into account using the BCS method with a density-independent pairing force. The strength parameters of the pairing force for neutrons and protons are $V_n = 349.5$ and $V_p = 330$~MeV$\cdot$fm$^3$, respectively~\cite{Zhao10}. The Dirac equation is solved iteratively starting from different initial Woods-Saxon potentials with deformation $\beta_2=-0.4, -0.2, 0.0, 0.2, 0.4$ for each nucleus until self-consistent convergence is achieved in order to pin down the ground state. We take the ground state corresponding to the mean-field state with the lowest energy.

The dynamical correlation effects associated with quadrupole deformations have been studied in the framework of projected GCM on top of a self-consistent nonrelativistic mean-field model using the SLy4 force~\cite{Bend05,Bend06}, where a topological Gaussian overlap approximation (GOA) for the exact AMP was used to reduce the computation time. Since the exact projected GCM based on the CDFT is time consuming and not feasible at this moment for the determination of the dynamical correlation energies of all even-even nuclei with available mass. Therefore, as the first step, we follow the same way as in Refs.~\cite{Gori05,Cham08} to calculate the dynamical correlation energies at the mean-field level, but on top of the CDFT.

The rotational correction energy $E_{\rm rot}$ associated with AMP is calculated with the cranking approximation,
\begin{equation}
   \label{eq1}
   E_{\rm rot}=\frac{\hbar^{2}}{2{\cal I}}\langle\hat{J}^{2}\rangle,
\end{equation}
where the moment of inertia ${\cal I}$ is calculated by the Inglis-Belyaev formula~\cite{Ingl56,Bely61} with $\hat{J}$ being the angular momentum operator. The full dynamical correlation energy, which is composed of both rotational and vibration correction energies (associated with GCM+AMP), can be parameterized by multiplying a deformation-dependent factor to the $E_{\rm rot}$~\cite{Gori05,Cham08},
\begin{equation}
\label{eq2}
E_{\rm corr}=E_{\rm rot}\{b{\rm tanh}(c|\beta_{2}|)+d|\beta_{2}|{\rm e}^{[-l(|\beta_{2}|-\beta_{2}^{0})^{2}]}\},
\end{equation}
with $\beta_2$ the calculated quadrupole deformation. The first term in the factor of Eq.(\ref{eq2}) is to correct the cranking formula in Eq.~(\ref{eq1}) for the rotational correction energy in order to obtain values close to those of an exact AMP calculation. The second term takes account of the deformation dependence of the vibrational correction, which vanishes for spherical nuclei. The parameters $b, c, d, l, \beta_2^0$ are respectively 0.80, 10, 2.6, 10, 0.10, as in Ref.~\cite{Cham08}. Actually, we have determined these parameters in our case by fitting the dynamical correlation energies to the mass difference between the experimental data and the pure mean-field results. The optimal parameters $b, c, d, l$ are respectively 1.0, 15, 1.6, 32, where $\beta_2^0$  has been fixed to be 0.10 in the fitting process as usual. We find that these two sets of parameters give the similar dynamical correlations energies for most nuclei, except for the heavier ones with $N>82$, where the new fitted parameters give the energy correction at most 0.4 MeV smaller than those by the old parameters of Ref.~\cite{Cham08}. However, which set of parameters give the dynamical correlations energies closer to the values by the exact projected GCM calculations is not clear. Moreover, the resultant rms deviation in the mass is quite close by the two sets of parameters. In view of these facts, we just take the same parameters as Ref.~\cite{Cham08} throughout this calculation.

The detailed results for the present calculation can be found in Ref.~\cite{Data}, which include the quadrupole deformation parameter $\beta_2$, total mean-field energy and rotational correction energy $E_{\rm rot}$ of the lowest mean-field state as well as the full dynamical correlation energy $E_{\rm corr}$ calculated by the prescription in Eq.~(\ref{eq2}). For spherical nuclei, as the cranking approximation calculation of rotational correction energy is ill-defined~\cite{Guzman00}, the corresponding rotational correction energies are imposed to be zero.

\begin{figure}[htb]
\includegraphics[width=8cm]{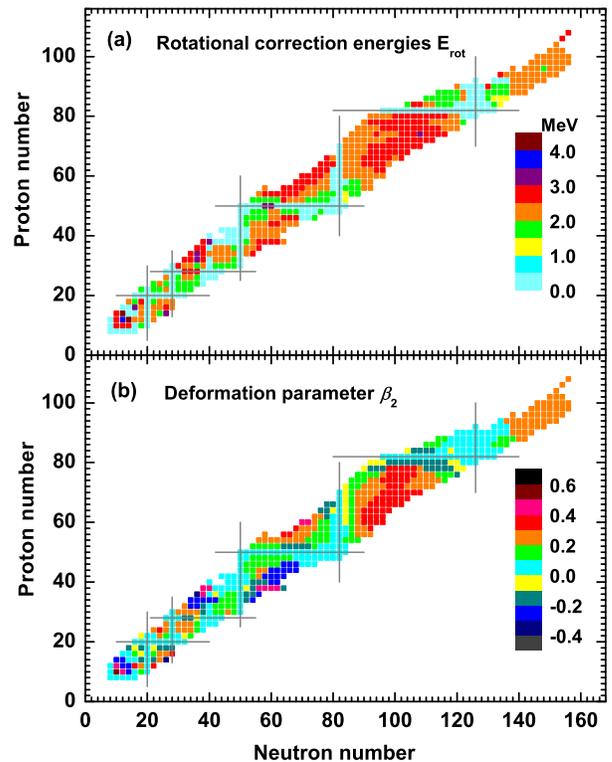}
\caption{\label{fig1}(Color online) Contour maps of the CDFT calculated ground state rotational correction energies $E_{\rm rot}$ and quadrupole deformation parameters $\beta_2$ by PC-PK1 as functions of the neutron and proton numbers. The rotational correction energies $E_{\rm rot}$ for spherical nuclei ($\beta_2=0$) are set to zero.}
\end{figure}

Figure~\ref{fig1} displays the contour maps of the CDFT calculated ground state rotational correction energies $E_{\rm rot}$ and quadrupole deformation parameters $\beta_2$ by PC-PK1~\cite{Zhao10} as functions of the neutron and proton numbers.  The main findings are as follows:
 \begin{enumerate}
 \item[(i)] The rotational correction energies vary between $\sim$1.5 MeV and $\sim$3 MeV for well-deformed nuclei. The correction energies in middle shell nuclei are significantly larger than those near the shell closure. For the mirror nuclei $^{26}$Mg and $^{26}$Si, the rotational correction energies are up to 4 MeV.
 \item[(ii)] Most of the nuclei are prolate deformed. There are some oblate deformed islands mainly located around subshell and shell closure with the proton and neutron numbers $(Z, N)\sim(14, 14)$, $(20, 28)$, $(34, 40)$, $(40, 60)$, $(50, 68)$, $(64,82)$, and $(82, 106)$, at which, an oblate-prolate shape transition or a shape-coexistence structure occurs~\cite{Li11,Heyde11}. For most of these nuclei, the corresponding rotational correction energies are $\sim$2.5 MeV.
\end{enumerate}

\begin{figure}[htb]
\includegraphics[width=8cm]{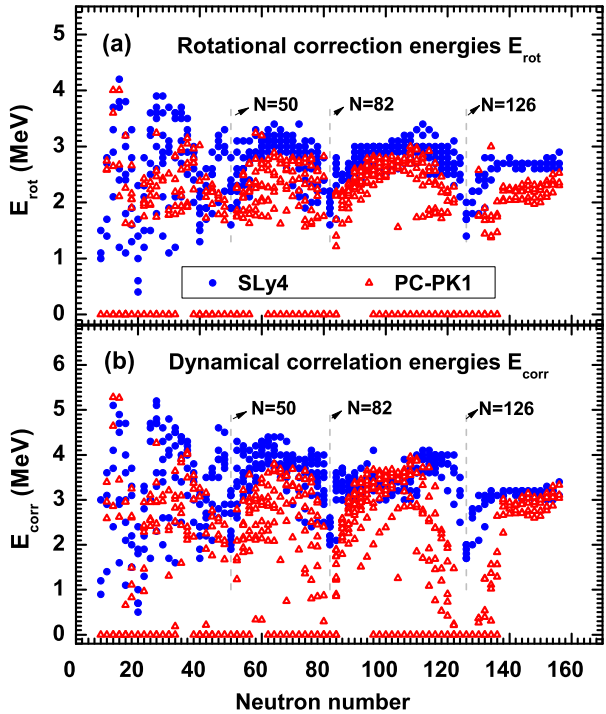}
\caption{\label{fig2}(Color online)
Rotational correction energies in Eq.~(\ref{eq1}) and full dynamical correlation energies in Eq.~(\ref{eq2}) calculated by CDFT with PC-PK1 (open triangles) in comparison with those~\cite{Bend06} from the pure AMP and the AMP+GCM (with the topological GOA) calculations by SLy4 (filled circles). In panel (a), the rotational correction energies $E_{\rm rot}$ from the cranking approximation calculation for spherical nuclei ($\beta_2=0$) are set to zero.}
\end{figure}

Figure~\ref{fig2} displays the rotational correction energies in Eq.~(\ref{eq1}) and full dynamical correlation energies in Eq.~(\ref{eq2}) calculated by CDFT with PC-PK1 (open triangles) in comparison with those~\cite{Bend06} from the pure AMP and AMP+GCM (with the topological GOA) calculations by SLy4 (filled circles). Again, the rotational correction energies for spherical nuclei ($\beta_2=0$) are assumed to be zero.

As shown in Fig.~\ref{fig2}, the systematic behaviors of both rotational and full dynamical correlation energies are similar in both calculations. However, the rotational correction energies from the cranking approximation are systematically smaller than those from the topological GOA of the AMP calculations~\cite{Bend06}. Excluding the spherical nuclei, the rms deviation of these two results is $\sim 0.64$~MeV. Furthermore, the full dynamical correlation energies are systematically underestimated by the prescription in Eq.~(\ref{eq2}), except for the nuclei in the middle of $N=82$ and $N=126$ shells, as well as for the nuclei beyond $N=126$, all of which are deformed heavy nuclei, cf. Fig.~\ref{fig1}. In comparison with the exact AMP calculation~\cite{Bend06}, it should be noted that the differences might be brought in by either the different treatments of rotational correction energies or the different energy density functionals. A global exact AMP calculation using the PC-PK1 force is needed to address this issue, which is beyond the scope of the present work.

\begin{figure}[htb]
\includegraphics[width=10cm]{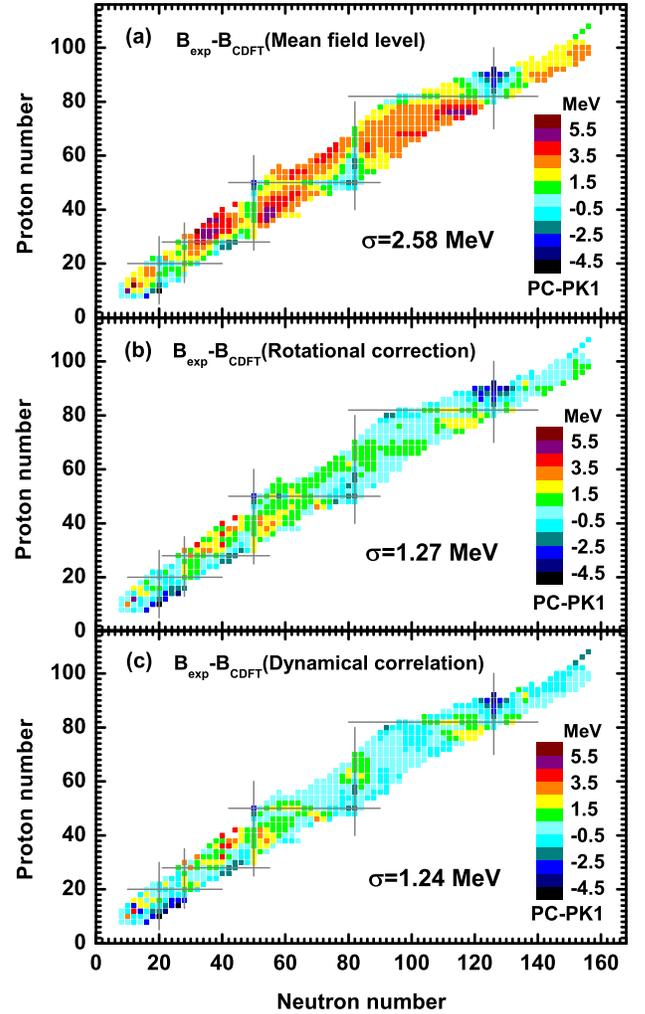}
\caption{\label{fig3}(Color online) Discrepancy of the CDFT calculated binding energies by PC-PK1 with the data for  575 even-even nuclei~\cite{audi03}. In panel (a), the CDFT calculated binding energies are given by the binding energies of the lowest mean-field states. In panels (b) and (c), the rotational correction energies and the full dynamical correlation energies are respectively taken into account.}
\end{figure}

Figure~\ref{fig3} displays the discrepancy of the CDFT calculated binding energies by PC-PK1 with the data for 575 even-even nuclei~\cite{audi03}. The main findings are as follows:
 \begin{enumerate}
 \item[(i)] In the CDFT calculation using PC-PK1, the binding energies are systematically underestimated by about 3 MeV. By including the rotational correction energies, which are in between 1.5 and 3 MeV as shown in Fig.~\ref{fig1}, the rms deviation is reduced from 2.58 MeV to 1.27 MeV.  The remaining discrepancy between the calculated and measured binding energies for most nuclei is in between -0.5 and 1.5 MeV. By including the full dynamical correlation energies, the rms deviation is reduced down to 1.24 MeV.

 \item[(ii)]  Figure~\ref{fig3}(c) shows that the binding energies for most nuclei with $Z>50$ are reproduced very well except for the nuclei around $Z=90$ and $N=126$, where the binding energies are already overestimated at the mean-field level. As the PC-PK1 force was adjusted with only 60 spherical nuclei, it is remarkable that the deviation of binding energies with the data is almost isospin-independent. This feature implies that the PC-PK1 is very promising in describing the masses of nuclei with extreme isospin values.

 \item[(iii)] For nuclei in the neighborhood of $(Z, N)\sim(40, 40)$, their binding energies are still underestimated by about 3.5 MeV. There are two possible reasons for these underestimations. One reason is that the ground states of these nuclei are predicted to be spherical at the mean-field level and therefore their dynamical correlation energies are imposed to be zero according to Eq.(\ref{eq2}). However, many experimental data indicate that these nuclei are deformed in the ground states. As discussed in many literatures~\cite{Rodriguez11,Fu13}, the spherical minimum in the deformation energy surface will shift to a deformed state when the dynamical correlation energies are taken into account. From this point of view, the rotational correction energies are actually not zero for these nuclei, e.g., for $^{76}$Kr, for which the dynamical correlation energy from an exact AMP+GCM calculation using the PC-PK1 force turns out to be 2.4 MeV~\cite{Fu13}. Another reason is due to the contribution from the Wigner energy in these nuclei with $N$ close to $Z$~\cite{Wigner37,Goriely02}. The inclusion of these effects will further improve the description of masses in this region. However, the detailed studies of all these effects are beyond the scope of the present study.

   \item[(iv)] For the nuclei that are located in the west of $(Z, N)\sim(44, 64), (78, 120)$ and in the east of $(Z, N)\sim(64, 78)$, the remaining discrepancies in the binding energies are expected to be reduced after inclusion nonaxial shape degrees of freedom, as demonstrated in the systematic finite-range liquid-drop model calculation~\cite{Moller06}, which has shown that there are islands of nonaxial shapes located around these regions. The correction energy from the nonaxial shape degrees of freedom could be up to 0.5 MeV.

 \item[(v)] The binding energies of the nuclei around $Z=14, 28, 90$ are not reproduced well. They are predicated to be overbound by $2-3$ MeV after including the full dynamical correlation energies. In Ref.~\cite{Geng05}, it has been discussed that the nuclei around $Z=90$ and $N=126$ are overestimated by RMF approaches. The spherical proton shell gap determined by the relative position of 1h$_{9/2}$ and 2f$_{7/2}$  orbitals plays an important role in reproducing the masses of nuclei in this region. The underlying reasons for the nuclei with $Z=14, 28$ are to be investigated in the future.

\end{enumerate}

In summary, we have carried out a global study of the dynamical correlation energies associated with the quadrupole shapes for the 575 even-even nuclei with proton numbers ranging from $Z=8$ to $Z=108$ within the CDFT using the PC-PK1 force. The rotational correction energy has been evaluated with the cranking approximation, while the vibrational correction was taken into account phenomenologically. We have compared our results with the values from the projected GCM (topological GOA) calculation using the SLy4 force. The systematic behavior of dynamical correlation energies is similar. However, our values from the cranking approximation calculation are overall smaller than those given by the projected GCM calculation. By including the dynamical correlation energies, the CDFT prediction with the PC-PK1 force for the masses of 575 even-even nuclei is improved with the rms deviation reduced from 2.58 MeV to 1.24 MeV. In particular, the deviation of binding energies with the data is almost isospin-independent. These features provide us confidence for its predictive power for $r$-process nuclei.

It should be pointed out that the present dynamical correlation energies are calculated with the wave functions of the mean-field ground states. For some nuclei such as $^{76}$Kr and $^{80}$Zr, the ground states will be altered in the beyond mean-field calculation, in which case,  one has to carry out an exact beyond mean-field calculation by configuration mixing of projected states. Work along this line in the framework of multi-reference CDFT is in progress.

\bigskip
We acknowledge S. Goriely, B. Sun, and P. W. Zhao for stimulating discussions. This work was supported in part by the National Undergraduate Training Programs for Innovation and Entrepreneurship (Project No.201210635132), the Major State 973 Program 2013CB834400, the NSFC under Grants No. 10975008, No. 10947013, No. 11175002, No. 11105110, No. 11105111, and No. 11205004, the Research Fund for the Doctoral Program of Higher Education under Grant No. 20110001110087, the Natural Science Foundation of Chongqing cstc2011jjA0376, and the Fundamental Research Funds for the Central Universities (XDJK2010B007 and XDJK2011B002).




\begin{thebibliography}{999}

\bibitem{tanihata85}I. Tanihata {\it et al.}, Phys. Rev. Lett. \textbf{55}, 2676 (1985).

 \bibitem{Mueller93} A. C. Mueller and B. M. Sherrill, Annu. Rev. Nucl. Part. Sci. \textbf{43}, 529 (1993).
 \bibitem{Tanihata95} I. Tanihata, Prog. Part. Nucl. Phys. \textbf{35}, 505 (1995).

\bibitem{meng96}J. Meng and P. Ring, Phys. Rev. Lett. \textbf{77}, 3963 (1996).

\bibitem{meng98}J. Meng and P. Ring, Phys. Rev. Lett. \textbf{80}, 460 (1998).

\bibitem{Sorlin08} O. Sorlin and M.-G. Porquet, Prog. Part. Nucl. Phys. \textbf{61}, 602 (2008).
\bibitem{Ozawa00} A. Ozawa, T. Kobayashi, T. Suzuki, K. Yoshida, and I. Tanihata, Phys. Rev. Lett. \textbf{84}, 5493 (2000).
\bibitem{Zhou10} S.-G. Zhou, J. Meng, P. Ring, and E.-G. Zhao, Phys. Rev. C \textbf{82}, 011301 (2010).
\bibitem{burbidge57}E. M. Burbidge, G. R. Burbidge, W. A. Fowler, F. Hoyle, Rev. Mod. Phys. \textbf{29}, 547 (1957).
\bibitem{Sun08} B. Sun, F. Montes, L. S. Geng, H. Geissel, Y. A. Litvinov, and J. Meng, Phys. Rev. C \textbf{78}, 025806 (2008).
\bibitem{Niu09} Z. M. Niu, B. Sun, and J. Meng, Phys. Rev. C \textbf{80}, 065806 (2009).
\bibitem{LiZ12} Z. Li, Z. M. Niu, B. Sun, N. Wang, and J. Meng, Acta Phys. Sin.-Ch. ED. \textbf{61}, 072601 (2012).
\bibitem{zhang12} W. H. Zhang, Z. M. Niu, F. Wang, X. B. Gong, and B. H. Sun, Acta Phys. Sin.-Ch. ED. \textbf{61}, 112601 (2012).
\bibitem{Xu13} X. D. Xu, B. Sun, Z. M. Niu, Z. Li, Y-Z. Qian, and J. Meng, Phys. Rev. C \textbf{87}, 015805 (2013).
\bibitem{Niu13} Z. M. Niu, Y. F. Niu, H. Z. Liang, W. H. Long, T. Nik\v{s}i\'{c}, D. Vretenar, and J. Meng, Phys. Lett. B \textbf{723}, 172 (2013).

 \bibitem{moller95}P. M\"{o}ller, J. R. Nix, W. D. Myers, and W. J. Swiatecki, At. Data Nucl. Data Tables \textbf{59}, 185 (1995).
 \bibitem{Bethe36} H. A. Bethe and R. F. Bacher, Rev. Mod. Phys. \textbf{8}, 82 (1936).
 \bibitem{Wang10}N. Wang, Z. Liang, M. Liu, and X. Wu, Phys. Rev. C \textbf{82}, 044304 (2010).
 \bibitem{Liu11} M. Liu, N. Wang, Y. Deng, and X. Wu, Phys. Rev. C \textbf{84}, 014333 (2011).
 \bibitem{DZ95} J. Duflo and A. P. Zuker, Phys. Rev. C \textbf{52}, R23 (1995).
 \bibitem{Koura05} H. Koura, T. Tachibana, M. Uno,and M. Yamada, Prog. Theor. Phys. \textbf{113}, 305 (2005).
 \bibitem{Bend05}M. Bender, G. F. Bertsch, and P.-H. Heenen, Phys. Rev. Lett. \textbf{94}, 102503 (2005).
 \bibitem{Bend06}M. Bender, G. F. Bertsch, and P.-H. Heenen, Phys. Rev. C \textbf{73}, 034322 (2006).
\bibitem{Gori09a} S. Goriely, N. Chamel, and J. M. Pearson, Phys. Rev. Lett. \textbf{102}, 152503 (2009).
\bibitem{Gori10} S. Goriely, N. Chamel, and J. M. Pearson, Phys. Rev. C \textbf{82}, 035804 (2010).

\bibitem{Gori09b}S. Goriely, S. Hilaire, M. Girod, and S. P\'{e}ru, Phys. Rev. Lett. \textbf{102}, 242501 (2009).

\bibitem{Reinhard89} P. G. Reinhard, Rep. Prog. Phys. \textbf{52}, 439 (1989).

\bibitem{Ring96} P. Ring, Prog. Part. Nucl. Phys. \textbf{37}, 193 (1996).

\bibitem{Vretenar05} D. Vretenar, A.~V. Afanasjev, G.~A. Lalazissis, and P. Ring, Phys. Rep. \textbf{409}, 101 (2005).

\bibitem{Meng06} J. Meng, H. Toki, S.-G. Zhou, S. Q. Zhang, W. H. Long, and L. S. Geng, 
                 Prog. Part. Nucl. Phys. \textbf{57}, 470 (2006).

\bibitem{Hirata97}D. Hirata, K. Sumiyoshi, I.Tanihata, Y. Sugahara, T. Tachibana, and H. Toki, Nucl. Phys. A \textbf{616}, 438 (1997).

\bibitem{Lalazissis99}G. A. Lalazissis, S. Raman, and P. Ring, At. Data Nucl. Data Tables \textbf{71}, 1 (1999).
\bibitem{Geng05} L. S. Geng, H. Toki, and J. Meng, Prog. Theor. Phys. \textbf{113}, 785 (2005).
\bibitem{Niks06}T. Nik\v{s}i\'{c}, D. Vretenar, and P. Ring, Phys. Rev. C \textbf{74}, 064309 (2006).
\bibitem{Yao08} J. M. Yao, J. Meng, P. Ring,  and D. Pena Arteaga, Chin. Phys. Lett. \textbf{25}, 3609 (2008).

\bibitem{Yao09}J. M. Yao, J. Meng, P. Ring, and D. Pena Arteaga, Phys. Rev. C \textbf{79}, 044312 (2009).

\bibitem{Yao10}J. M. Yao, J. Meng, P. Ring, and D. Vretenar, Phys. Rev. C \textbf{81}, 044311 (2010).

\bibitem{Niks09}T. Nik\v{s}i\'{c}, Z. P. Li, D. Vretenar, L. Pr\'{o}chniak, J. Meng, and P. Ring, 
                Phys. Rev. C \textbf{79}, 034303 (2009).

\bibitem{Li09} Z. P. Li, T. Nik\v{s}i\'{c}, D.Vretenar, J. Meng, G. A. Lalazissis, and P. Ring, 
               Phys. Rev. C \textbf{79}, 054301 (2009).

\bibitem{Niu2013PRC} Z. M. Niu, Y. F. Niu, Q. Liu, H. Z. Liang, and J. Y. Guo, Phys. Rev C \textbf{87}, 051303(R) (2013).

\bibitem{Yao11} J. M. Yao, H. Mei, H. Chen, J. Meng, P. Ring, and D. Vretenar, Phys. Rev. C \textbf{83}, 014308 (2011).

\bibitem{Zhao10}P. W. Zhao, Z. P. Li, J. M. Yao, and J. Meng, Phys. Rev. C \textbf{82}, 054319 (2010).

\bibitem{Zhao11a} P. W. Zhao, S. Q. Zhang, J. Peng, H. Z. Liang, P. Ring, and J. Meng, Phys. Lett. B \textbf{699}, 181 (2011).

\bibitem{Zhao11b} P. W. Zhao, J. Peng, H. Z. Liang, P. Ring, and J. Meng, Phys. Rev. Lett. \textbf{107}, 122501 (2011).

\bibitem{Xiang12} J. Xiang, Z. P. Li, Z. X. Li, J. M. Yao, and J. Meng, Nucl. Phys. A \textbf{873}, 1 (2012).

\bibitem{Li12} Z. P. Li, C. Y. Li, J. Xiang, J. M. Yao, and J. Meng, Phys. Lett. B \textbf{717}, 470 (2012).

\bibitem{hua12} X. M. Hua, T. H. Heng, Z. M. Niu, B. Sun, and J. Y. Guo, Sci. China Phys. Mech. Astron. \textbf{55}, 2414 (2012)

\bibitem{Zhao12} P. W. Zhao, L. S. Song, B. Sun, H. Geissel, and J. Meng, Phys. Rev. C \textbf{86}, 064324 (2012).

\bibitem{Meng13} J. Meng, J. Peng, S. Q. Zhang, and P. W. Zhao, Frontiers of Physics, \textbf{8}, 55 (2013).


\bibitem{Qu13} X. Y. Qu,  Y. Chen,  S. Q. Zhang,  P. W. Zhao,  I. J. Shin,  Y. Lim,  Y. Kim, and J. Meng,
              Sci. China Phys. Mech. Astron. \textbf{56}, 2031 (2013).

\bibitem{Burv02} T. B\"urvenich, D. G. Madland, J. A. Maruhn, and P.-G. Reinhard, Phys. Rev. C \textbf{65}, 044308 (2002).

\bibitem{Gori05} S. Goriely, M. Samyn, J. M. Pearson, and M. Onsi, Nucl. Phys. A \textbf{750}, 425 (2005).

\bibitem{Cham08}N. Chamel, S. Goriely, and J. M. Pearson, Nucl. Phys. A \textbf{812}, 72 (2008).

\bibitem{Ingl56}D. Inglis, Phys. Rev. \textbf{103}, 1786 (1956).

\bibitem{Bely61}S. Belyaev, Nucl. Phys. A \textbf{24}, 322 (1961).


\bibitem{Data}  See Supplemental files for the detailed results.

\bibitem{Guzman00} R. R. Rodr\'{\i}guez-Guzm\'{a}n, J. L. Egido, and L. M. Robledo, Phys. Lett. B \textbf{474}, 15 (2000).

\bibitem{Li11} Z. P. Li, J. M. Yao, D. Vretenar, T. Nik\v{s}i\'{c}, H. Chen, and J. Meng, Phys. Rev. C \textbf{84}, 054304 (2011).

 \bibitem{Heyde11}   K. Heyde and J. L. Wood, Rev. Mod. Phys. \textbf{83}, 1467 (2011).

\bibitem{audi03} G. Audi, A. H. Wapstra, and C. Thibault, Nucl. Phys. A \textbf{729}, 337 (2003).

\bibitem{Rodriguez11}T. R. Rodr\'{\i}guez and J. L. Egido, Phys. Lett. B \textbf{705}, 255 (2011)

\bibitem{Fu13}   Y. Fu, H. Mei, J. Xiang, Z. P. Li, J. M. Yao, and J. Meng,
                   Phys. Rev. C \textbf{87}, 054305 (2013).
 \bibitem{Wigner37} E. Wigner, Phys. Rev. \textbf{51}, 106 (1937).
 \bibitem{Goriely02}S. Goriely, M. Samyn, P.-H. Heenen, J. M. Pearson, and F. Tondeur,
                    Phys. Rev. C \textbf{66}, 024326 (2002).
 \bibitem{Moller06} P. M\"{o}ller, R. Bengtsson, B. G. Carlsson, P. Olivius, and T. Ichikawa,
                    Phys. Rev. Lett. \textbf{97}, 162502 (2006).

\end{thebibliography}
\end{document}